\documentclass[12pt,fleqn]{article}

\usepackage{amsmath, amssymb, verbatim}
\usepackage[dvips]{graphicx}
\newcommand \be{\begin{equation}}
\newcommand \ee{\end{equation}}
\newcommand \bea{\begin{eqnarray}}
\newcommand \eea{\end{eqnarray}}
\newcommand \bee{\begin{equation}}
\newcommand \eee{\end{equation}}

\newcommand\TL{\hfil$\displaystyle{##}$}
\newcommand\TR{$\displaystyle{{}##}$\hfil}
\newcommand\TC{\hfil$\displaystyle{##}$\hfil}

\def\seqalign#1#2{\vcenter{\openup1\jot
  \halign{\strut #1\cr #2 \cr}}}
\def\lbldef#1#2{\expandafter\gdef\csname #1\endcsname {#2}}
\newcommand{\eqn}[3][]{\lbldef{#2}{(\ref{#2})}%
\def\@eqnstyle{#1}%
\ifx\@eqnstyle\@empty%
\begin{equation} \eqalign{#3} \label{#2} \end{equation}%
\else%
\begin{equation} \seqalign{\span\TC}{#3} \label{#2} \end{equation}%
\fi}
\def\eqalign#1{\vcenter{\openup1\jot
    \halign{\strut\span\TL & \span\TR\cr #1 \cr
   }}}

\def\p2{{p \over 2}}

\def\nref#1{(\ref{#1})}

\input epsf

\begin{document}


\thispagestyle{empty}
\renewcommand{\thefootnote}{\fnsymbol{footnote}}

{\hfill \parbox{4cm}{ arXiv: 0907.1625 \\
  NSF-KITP-09-115\\
  }}

\bigskip\bigskip

\begin{center} \noindent \Large \bf
Higher Derivative Gravity, Causality and Positivity of Energy in a UV complete QFT
\end{center}

\bigskip\bigskip\bigskip

\centerline{ \normalsize \bf Diego  M. Hofman$^a$\footnote[1]{\noindent \tt dhofman@Princeton.edu} }

\bigskip
\bigskip\bigskip

\centerline{$^a$ \it Joseph Henry Laboratories, Princeton
University, Princeton, NJ 08544, USA}
\bigskip

\bigskip\bigskip

\renewcommand{\thefootnote}{\arabic{footnote}}

\centerline{\bf \small Abstract}
\medskip

{\small In this note we discuss the relation between the constraints imposed by causality in the bulk of $AdS$ and the condition of
positivity of the energy measured in ideal calorimeters in a collider experiment in the dual CFT. We first extend the analysis in the literature and  recover all bounds imposed by causality of the boundary theory in the bulk dynamics for all
polarizations of the graviton and the gauge boson field. These results translate to specific bounds for the ratio
of central charges $\frac{a}{c}$ in the dual CFT, already found by analyzing the energy one point function. Then, we generalize this discussion and we study shock wave backgrounds in which we
make manifest the relation between causality in the bulk and the three point function in the dual field theory. We remark that particular care
has to be given to the exponentiation procedure of the three point function when solving the classical equations of motion in the higher gravity
theory, as it is not clear that every theory will present causality problems. Finally, we present a field theoretic argument explaining the positivity of energy condition in any UV complete QFT. 

 }

\newpage

\section{Introduction}

In the last few years we have seen a great deal of progress in understanding fundamental properties of quantum field theories and quantum
gravity through the AdS/CFT correspondence \cite{adscft,adswitten,adsgkp}. Because this is a weak-strong coupling duality, it is a common
characteristic of the correspondence that some features of the theory under consideration may appear in different guises as we study different
regimes. While we have a weak coupling description in terms of a local quantum field theory,  on the other hand we study the strong coupling
regime as a classical theory of gravity (in the strong 't hooft coupling limit). It is clear that in these two extreme cases the fundamental degrees of freedom
of the theory become reorganized and we obtain very different descriptions of the same phenomena. This is one of the reasons the AdS/CFT
correspondence is so powerful.


The main phenomenon we would like to discuss in this note was discovered from two different perspectives. On the one side, one of the most
interesting new insights discovered through gauge/gravity duality is the presence of a universal behavior for the ratio of the shear viscosity
($\eta$) and entropy ($s$) density for field theories with an Einstein gravity dual. Because at strong coupling (when one would expect this
ratio to be lower for a theory, as the mean free path goes to zero) all theories with an Einstein gravity dual converge to the same value of
$\frac{\eta}{s}$, namely, $\frac{1}{4 \pi}$, it was proposed, early on, that $\frac{\eta}{s}$ could be bound from below (KSS) \cite{kss}. This
observation led the authors of \cite{kats} and \cite{liu} to consider what would happen if one considered higher derivative corrections to
Einstein gravity\footnote{In \cite{Buchel:2008vz}, gauge theories with large central charges were studied where these higher derivative terms are present in the dual gravitational description. For certain values of the central charges the KSS bound was found to be violated.}. It was understood in these papers that, unless some bound was imposed on the coefficients in the action for the higher
derivative terms, the KSS bound would be violated. It was suggested in \cite{liu} that requiring the theory on the boundary to be causal imposed
such a bound\footnote{Previously, it had been noticed in \cite{Cvetic:2001bk} that a less restrictive bound could be obtained by requiring that the entropy of black holes be positive.}. This restriction was insufficient to preserve the KSS bound. The new bound obtained in \cite{liu} for a theory which is dual to five
dimensional Gauss-Bonnet (GB) gravity  was $\frac{\eta}{s} \geq \frac{16}{25}\frac{1}{4\pi}$ \cite{liu}. It is now understood that the KSS bound
can't be correct and it is not clear whether any other bound could exist at all.

The action of five dimensional
GB gravity is:

\begin{equation}\label{gbaction}
S_{GB} = \frac{1}{16 \pi G_N} \int dx^5 \sqrt{g}\left[ R + 12 + \frac{\lambda}{2} \left(R^2- 4 R_{\mu\nu}R^{\mu\nu} + R_{\mu\nu\rho\sigma}R^{\mu\nu\rho\sigma} \right)\right]
\end{equation}

\noindent where the $AdS$ radius of the $\lambda=0$ solution was normalized to 1. This theory is special in the sense that it is the only
one among 4 derivative actions of gravity in five dimensions in which the equations of motion for a perturbation propagating in a given background
have only 2 derivatives. This makes the classical propagation of perturbations a well defined initial conditions problem.  The causality bound  in \cite{liu} was obtained by propagating a helicity
2 graviton state\footnote{We define helicity as the eigenvalue of the state under $SO(2)$ rotations in the transverse coordinates. Notice that the theory also possesses helicity 1 and helicity 0 degrees of freedom as the 5 dimensional
massless graviton multiplet has 5 excitations. We will make use of this fact shortly.}  in a black hole background of Gauss-Bonnet gravity and
noticing that if such a state were to be ``dropped'' from the boundary it would bounce back to it and land outside the light cone of the dual
field theory. Therefore, causality in the boundary implied a bound on the coefficient of Gauss-Bonnet gravity in the bulk, $\lambda \leq
\frac{9}{100}$. The upshot of this discussion is that certain restrictions of consistency in the field theory constrain the dual gravity action.

Having discussed the constraints in the gravitational theory, what is the easiest way to study the same physics in the field theory?

The other road that led to the discovery of equivalent physical bounds was the study of collider physics of conformal theories in
\cite{ccp}. With the era of the LHC around the corner it makes sense to ask what insights can we get from AdS/CFT into collider physics. This is
particularly interesting for (beyond the Standard) Models where there there is a hidden conformal sector, as in \cite{rs,
Georgi:2007ek,Strassler:2006im}. This type of systematic study was undertaken in \cite{ccp} where the most general form of the energy and charge
correlation functions \cite{energy} were calculated for a CFT. Because any UV complete theory is either conformal or free, this analysis
describes the most general behavior in field theory at short distances\footnote{The behavior of the theory at long distances can certainly change the readings at calorimeters in a collider experiment. This happens in a similar fashion to the way hadronization affects high energy QCD results.}. The energy one point function determines the expected value of the
energy accumulated in a calorimeter at the end of a collider experiment. Therefore, one expects this quantity to be positive. If the state in
which we calculate the expectation value is created by the energy-momentum tensor and we have $\mathcal{N}=1$ supersymmetry, then, the
correlation function depends only on the central charges of the theory. Furthermore, the same is true for states created by $R$ currents, which
are in the same supermultiplet. It turns out, as remarked in \cite{ccp}, that one of the bounds obtained from the requirement that the energy
deposited in calorimeters is positive coincides, for $\mathcal{N}=1$ theories, with the bound discussed above in Gauss-Bonnet gravity. It is of
great interest to note that while energy correlation functions are observables that were introduced because of their relevance to experimental
setups, they seem to give us some new insights on the fundamental structure of our theories, imposing physical constraints. 

We will discuss how these two approaches are related. It is a fact that we should look at \nref{gbaction} as an effective
action specifically engineered to reproduce the two and three point functions of the energy-momentum tensor in the dual field theory. All other
information contained in \nref{gbaction} should not be trusted at $\lambda$ of order 1. The knowledge of two and three point functions amounts to knowing the
energy one point function, which is the observable we discuss in the CFT. In general, classical calculations using higher derivative gravity actions involve more information than two and three point functions. We will show that peculiarities of GB gravity imply that the causality problem is directly connected with the energy one point function alone. Incidentally, there are other ways to parameterize the 4 derivative
gravity action, such that the energy one point function is reproduced. We will discuss them below, but it is important to stress that they are
not equivalent to GB gravity at the nonlinear level. Here, as the exponentiation procedure in the classical solutions is more involved, it is not clear whether there are any causality problems of the type discussed for GB gravity.

This paper is organized in the following way. In section \ref{resultsccp} we present the necessary results from \cite{ccp}  and we add a few new expressions concerning energy
correlation functions and the higher derivative gravity theories that reproduce these results.  In section \ref{black} we recall and extend the
analysis in \cite{liu} for black hole backgrounds and recover all bounds discussed in \cite{ccp} in a systematic fashion. In section \ref{shock} we discuss scattering
gravitational and gauge field perturbations of a gravitational shock wave. This discussion makes manifest the connection of the gravity
calculation with the energy one point function in the CFT.  We also discuss the
effects of considering different higher derivative actions instead of Gauss-Bonnet gravity.
 In section \ref{positivity} we present an argument in favor of the positivity of energy
condition in any CFT. This argument is completely general and applies to strongly coupled as well as non supersymmetric theories.
We conclude in section \ref{disc} where we present a discussion of the results.

\section{Some results for energy one point functions}
\label{resultsccp}

In this section we recall some results concerning energy one point functions in CFTs and the dual gravitational actions that reproduce them. We
follow the discussion in \cite{ccp} closely but stress the results important for the problem at hand.

The $n$ point energy correlation function is defined as

\begin{equation}\label{npoint}
\langle {\cal E}(\theta_1) \cdots {\cal E}(\theta_n) \rangle_\mathcal{O} \equiv  \frac{ \langle 0 | {\cal O}^\dagger
 {\cal E}(\theta_1) \cdots {\cal E}(\theta_n) { \cal O} |0 \rangle }{
 \langle 0 | {\cal O}^\dagger
   { \cal O} |0 \rangle}
   \end{equation}

\noindent where $\mathcal{O}$ is an operator that creates an initial state and the energy operator $\mathcal{E}$ is given by

\begin{equation}\label{eop}
 {\cal E}(\vec{n}) = \lim _{r \to \infty} r^2  \int_{-\infty}^\infty dt \, n^i T^0_{~ i}(t, r \vec n^i)
\end{equation}

We point out that, for states created by local operators, energy $n$ point functions are connected to $n+2$ correlation functions in the field
theory: $n$ energy momentum tensors and 2 extra insertions given by the operator $\mathcal{O}$ (which could be the energy momentum tensor as well).

It is easy to see that the expectation value corresponding to the one point function represents the total energy deposited in a calorimeter at a
solid angle $\theta$ at the end of an experiment where the initial state is $ \mathcal{O} |0 \rangle $. As it is clear from expression
\nref{npoint}, the energy one point function can be obtained from integration over the three point function of local operators (if $\mathcal{O}$
is local) and the energy-momentum tensor. Because 3 point functions are fixed in a conformal theory, except for a finite number of parameters
\cite{Osborn:1993cr}, we can calculate the most general form of the energy one point function. It makes sense to consider as operators
$\mathcal{O}$ a conserved current $J^\mu$ or the energy-momentum tensor itself $T^{\mu\nu}$ as these are generally present in arbitrary
theories. The most general form for this one point functions is

\begin{eqnarray}
\langle {\cal E }(\vec{n}) \rangle_{J^i\epsilon_i} & = &   {  q \over  4 \pi }
  \left[  1  + a_2
  ( \cos^2 \theta  - { 1 \over 3 }  ) \right]\\
  \langle {\cal E}(\vec{n}) \rangle_{T^{ij}\epsilon_{ij}} & = & { q \over 4 \pi }
    \left[
   1 + t_2
   \left( { \epsilon_{ij}^* \epsilon_{il} n_i n_j \over   \epsilon_{ij}^* \epsilon_{ij}} - { 1 \over 3 }
   \right)
 + t_4
 \left( {|\epsilon_{ij} n_i n_j |^2 \over   \epsilon_{ij}^* \epsilon_{ij}} - { 2 \over 15 } \right)
  \right]
\end{eqnarray}

\noindent where $\theta$ is the angle between $\vec{n}$ and the (space-like) polarization of the current $\epsilon^i$ and $q$ is the total
energy of the state; $\epsilon_{ij}$ is the polarization tensor of the state created by the energy-momentum tensor. Although this form is fixed
in a CFT, the parameters $a_2$, $t_2$ and $t_4$ are not fixed by symmetries. There is, however, a set of constraints that has to be satisfied if
one demands that calorimeters can only pick up positive energies. These are:

\begin{eqnarray}
3-a_2 & \geq & 0\label{a21}\\
a_2 + \frac{3}{2} & \geq & 0 \label{a22}\\
( 1 - \frac{ t_2}{ 3} - \frac{ 2  t_4}{15 }) & \geq & 0\label{t21}\\
 2 ( 1 - \frac{ t_2 }{ 3} - \frac{ 2  t_4 }{15} ) + t_2 & \geq & 0\label{t22}\\
  \frac{3}{2 } ( 1 - \frac{ t_2}{3} - \frac{ 2  t_4}{15} ) + t_2 + t_4 & \geq & 0\label{t23}
  \end{eqnarray}

It is very interesting to note that the three conditions on $t_2$ and $t_4$ given in equations \nref{t21},\nref{t22} and \nref{t23} come,
respectively, from the helicity 2, 1 and 0 components of the polarization tensor $\epsilon_{ij}$ with respect to $SO(2)$ around $\vec{n}$, while
equations \nref{a21} and \nref{a22} come from the helicity 1 and 0 components of $\epsilon_i$. For reasons that will be clear momentarily we
will be interested in theories with at least $\mathcal{N}=1$ supersymmetry. In that case it is possible to write the coefficients $a_2$, $t_2$
and $t_4$ as a function of the central charges of the the theory, $a$ and $c$. These charges are defined by computing the trace anomaly as

\begin{equation}
T^\mu_\mu = \frac{ c}{ 16 \pi^2 } W_{\mu\nu\delta\sigma} W^{\mu\nu\delta\sigma} - \frac{  a}{16 \pi^2 } \left( R_{\mu \nu \delta \rho}R^{\mu \nu
\delta \rho} - 4 R_{\mu \nu} R^{\mu \nu} + R^2 \right)
\end{equation}

\noindent where $W$ is the Weyl tensor and $R$ represents the curvature tensors of the background. Imposing $\mathcal{N}=1$ supersymmetry
implies \cite{ccp}

\begin{equation}
t_2  =  6 \left(1 -\frac{a}{c}\right) \quad\quad ; \quad\quad t_4 =0 \label{t2t4}
\end{equation}

while

\begin{eqnarray}
a_2  =  0 \quad \quad & & \textrm{If} \, J^\mu \, \textrm{is a non R current}  \\
a_2^{U(1)} =  3 \left(1 -\frac{a}{c}\right) \quad \quad & & \textrm{If} \,
 J^\mu \, \textrm{ is a U(1) R current} \label{u1}
\end{eqnarray}

These results have been obtained in \cite{ccp} in the following manner. We first use that, for free theories,

\begin{equation}
 a_2^{free} = { 3  } \frac{  \sum_i (q_i^b)^2 - (q_i^{wf})^2}{
 \sum_i (q_i^b)^2 +  2 (q_i^{wf})^2  }
\end{equation}

 \noindent where we sum over the charges of all complex bosons and Weyl fermions for a given global symmetry. Then the result can be expressed
 as a function of $a$ and $c$ and, thus, has to be valid for all conformal theories, whether free or not.

 We can repeat the same calculation here for a theory with $\mathcal{N}=2$ symmetry
 where there is an $SU(2)$ R current. In that case the result is

\begin{equation}
a_2^{SU(2)} =  6 \left(1 -\frac{a}{c}\right) \label{su2}
\end{equation}

Using formulae \nref{t2t4}, \nref{u1} and \nref{su2} in inequalities \nref{a21}-\nref{t23}, coming from requiring positivity of the energy one point function in different states, we obtain

\begin{equation}\label{boundsac}
\frac{a}{c} \quad \left\{
  \begin{array}{ll}
    \geq 0, & \hbox{for U(1) R current, helicity 1 state;} \\
    \leq \frac{3}{2}, & \hbox{for U(1) R current, helicity 0 state;} \\
    \geq \frac{1}{2}, & \hbox{for SU(2) R current, helicity 1 state;} \\
    \leq \frac{5}{4}, & \hbox{for SU(2) R current, helicity 0 state;} \\
    \geq \frac{1}{2}, & \hbox{for energy-momentum, helicity 2 state;} \\
    \leq 2, & \hbox{for energy-momentum, helicity 1 state;} \\
    \leq \frac{3}{2}, & \hbox{for energy-momentum, helicity 0 state.}
  \end{array}
\right.
\end{equation}

Putting together this information we obtain the bounds in \cite{ccp}

\begin{eqnarray}
\frac{3}{2} \geq & \frac{a}{c} & \geq \frac{1}{2} \quad\quad \textrm{for} \,\, \mathcal{N}=1\\
\frac{5}{4} \geq & \frac{a}{c} & \geq \frac{1}{2} \quad\quad \textrm{for} \,\, \mathcal{N}=2
\end{eqnarray}

We would now like to discuss the gravitational higher derivative actions that reproduce these results by using the AdS/CFT dictionary. We
parameterize the actions by using field redefinitions and only pay attention to reproducing the exact energy one point functions. In that case
the general form of the action is \cite{ccp}:

\begin{equation}\label{gaction}
S = \frac{ M_{pl}^3 }{ 2 } \left[ \int d^5 x \sqrt{g} \left(R + 12 + \frac{ t_2}{48} W_{\mu \nu \delta \sigma }
   W^{\mu \nu \delta \sigma}\right)    \right] - \frac{ 1}{ 4 g^2 }\left[ \int  d^5 x \sqrt{g} \left( F^2 +
  \frac{a_2 }{12}  
  W^{\mu\nu \delta \rho} F_{\mu \nu} F_{\delta \rho}\right)\right]
\end{equation}

\noindent where we have normalized the radius of the $AdS$ solution to 1. In writing \nref{gaction} we have assumed $t_4=0$, otherwise we would have had $R^3$ contributions. This is the main reason we choose to
consider supersymmetric theories, as their gravity duals can be restricted to 4 derivatives actions, as far as the energy one point function is
considered. Notice that we have only included one generic field strength field in \nref{gaction}. When it comes to the fields that couple to the
R currents of the dual field theory we need to consider a $U(1)$ gauge field when there is $\mathcal{N}=1$ supersymmetry at the boundary. We
also must include an $SU(2)$ gauge field if there is $\mathcal{N}=2$ supersymmetry. The corresponding coefficients $a_2$ are given by \nref{u1}
and \nref{su2}.

The action \nref{gaction} is only to be understood as a \textit{machine} to reproduce the exact energy one point functions (three point functions of local
operators). Then, it is alright to consider coefficients $a_2$ and $t_2$ of order unity while ignoring higher derivative terms. It is not correct to take \nref{gaction} as full
nonlinear general description. A reflection of this fact is that \nref{gaction} posses ghost modes that violate unitarity semiclassically.
These modes are present for generic values of the couplings and are not connected to the causality problems discussed before in the
introduction. In fact, they should be \textit{canceled} in the full string theory description.

  It
is important to stress that although this action is exact in $a_2$, it might have higher order correction in $t_2$. As a matter of fact, we can
improve this result to the exact expression. In order to do that, we use the results in \cite{Nojiri:1999mh} where the central charges $a$ and
$c$ were calculated for the dual theories to higher derivative gravity actions. From \nref{t2t4} we can express $t_2$ as a function of $a$ and
$c$ and obtain an exact expression for the action, up to four derivative terms. The result is

\begin{equation}\label{gaction2}
S = \frac{ M_{pl}^3 }{ 2 } \left[ \int d^5 x \sqrt{g}\left( R + 12+ \frac{ t_2}{48 - 8 t_2} W_{\mu \nu \delta \sigma }
   W^{\mu \nu \delta \sigma} \right)   \right] - \frac{ 1}{ 4 g^2 }\left[ \int  d^5 x \sqrt{g} \left(  F^2 +
  \frac{a_2 }{12 }  
  W^{\mu\nu \delta \rho} F_{\mu \nu} F_{\delta \rho}\right)\right]
\end{equation}

Another important point is that other parameterizations are possible (although not completely equivalent as we will discuss below). The
authors of \cite{liu} have considered a GB term \nref{gbaction} instead of the $W^2$ term in \nref{gaction2}.
In that case the map between $\frac{a}{c}$ and $\lambda$ in \nref{gbaction} is \cite{Nojiri:1999mh,liu}:

\begin{equation}\label{acgb}
\frac{a}{c} = 3 - \frac{2}{\sqrt{1-4\lambda}}
\end{equation}

We are now in a position to use actions \nref{gaction2} and \nref{gbaction} to discuss how bounds \nref{a21}-\nref{t23} come about.

\section{Black hole backgrounds in higher derivative gravity}
\label{black}

It was argued in \cite{liu} that Gauss Bonnet gravity, for certain values of the coupling constant, is such that propagation of gravitons in the bulk lead to causality violations in the boundary dual theory. The setup is such that we have a black hole in the bulk of AdS. Then, it happens that we can
shoot particles from the boundary to the black hole in such a way that they bounce back to the boundary and end up traveling faster that
massless particles in the boundary. The argument uses that the velocity in a direction parallel to the boundary is greater than 1 in the bulk.
Because the effective black hole potential has a maximum, semiclassical (WKB) particles will stay there for a long time, making their average
velocity equal to the velocity at the maximum. The authors of \cite{liu} showed that the GB coupling constant $\lambda$ must be $\lambda \leq
\frac{9}{100}$ if we are to avoid this behavior. The black hole metric for Gauss Bonnet gravity is:

\begin{equation}
ds^2 = -  \frac{1}{2}\left(1+\sqrt{1-4\lambda}\right) f(r) dt^2 + \frac{1}{f(r)} dr^2 + r^2 dx^i dx^i
\end{equation}

\noindent where

\begin{equation}
f(r) = \frac{r^2}{2 \lambda} \left(1-\sqrt{1-4\lambda + \frac{4 \lambda}{r^4}}\right)
\end{equation}

\noindent and $i=1,2,3$. The study was performed for gravitons of helicity 2 (i.e. $h=h_{12}=h_{21}$ for a state that
propagates in the $3$ direction in the boundary and also inside the bulk (in the $r$ direction)). Authors of \cite{liu} were interested in a bound that would
constrain positive values of $\lambda$ as the viscosity to entropy ratio falls below $\frac{1}{4\pi}$ for $\lambda>0$. They did not study,
however, the possibility of a lower bound for $\lambda$. We would like to study the causality constraints more systematically and make a connection with the energy one point function argument coming from the boundary theory.

It turns out that we can map the bound on $\lambda$ to a bound on $\frac{a}{c}$ in the dual CFT by using \nref{acgb}, $\frac{a}{c} = 3 - \frac{2}{\sqrt{1-4\lambda}}$. If $\lambda > \frac{1}{4}$ the theory does not admit an $AdS$ solution. Therefore, we assume
$\lambda<\frac{1}{4}$. In this range $\frac{a}{c}$ is monotonous in $\lambda$. Then, the bound $\lambda<\frac{9}{100}$ implies
$\frac{a}{c}>\frac{1}{2}$.

It is interesting that this bound matches the lower bound of $\frac{a}{c}$ obtained in \cite{ccp} by demanding that the expectation value of the
energy measured in a calorimeter at some angle from some operator insertion is positive in a CFT. Notice that this bound matches the expected result for the helicity 2 component of the energy-momentum tensor in the dual boundary theory, \nref{boundsac}. Therefore, we should expect similar bounds coming from the analysis of other helicities of the graviton. In particular, the other optimal bound should come from the helicity 0 component. If we use \nref{acgb} we should expect a causality problem unless $\lambda > -\frac{7}{36}$.

Before showing that this is actually the case let us make another comment. An interesting remark is that the way the bound was obtained originally in \cite{liu} makes it hard to see that this feature is connected to the  UV properties of the boundary field theory. The reason is that the particle must explore deep into the bulk of $AdS$ to violate causality with the arguments presented in \cite{liu}.

One could try to calculate the exact geodesics in the effective GB metric for particles that don't go too deep into AdS. Instead we could try to consider a different background given by a shock wave. This corresponds to a localized insertion in
the CFT and is the actual dual of the energy one point function, as discussed in \cite{ccp}. Hopefully, we won't have to explore too deep into the bulk to see a problem, meaning this is related to UV properties of the field theory. Another motivation to do this is that the black hole
corresponds to heating up the CFT to a certain temperature. We would like to understand the CFT at zero temperature, instead.

In the remainder of this section we will discuss several computations in the black hole background while in the following section we discuss the shock wave approach.

\subsection{Bounds on $\frac{a}{c}$ from causality for all graviton polarizations}
\label{gravbh}

The basic object we would like to study, following \cite{liu},  is $c^2(r)$, the square of the speed of light in the 3-direction as a function of the $AdS$ coordinate\footnote{This
coordinate is such that the horizon of the black hole is at $r=1$ and the boundary of $AdS$ is at $r=\infty$.} $r$ for a given perturbation of the metric, $h_{\mu\nu}$. Because a massless particle does not follow geodesics in Gauss Bonnet gravity, the way to calculate this is to perturb the black hole
metric by $ h_{\mu\nu}(r,x^0,x^3) dx^\mu dx^\nu$ and obtain the quadratic effective action for it. After appropriate gauge fixing and separation of the modes, this will result in an action of the type of a scalar in
a non trivial metric. We can read off the components of the metric from the effective action as:

\begin{equation}
S_{eff}[h] = \int d^5x g^{rr} \partial_r h \partial_r h + g^{33} \partial_3 h \partial_3 h + g^{00} \partial_0 h \partial_0 h + \ldots
\end{equation}

From this action we can read the speed of light in the 3-direction:

\begin{equation}
c^2 =\left(\frac{dx^3}{dx^0}\right)^2 = - \frac{g^{00}}{g^{33}}
\end{equation}

Alternatively we could look at the  equation of motion for the perturbation.

Because $c^2$ is the effective potential for our semiclassical particle \cite{liu}, one possible approach is to look for a maximum of $c^2(r)$ such that $c^2(r)>1$. Then we can tune the probe particle to spend most of its trajectory here at speed $\sqrt{c^2_{max}}$ and then bounce back to the boundary. In that case we will have a violation of causality as the average speed of the particle will be greater than 1.

The result for the helicity 2 mode $h_{12}$ was obtained in \cite{liu}

\begin{equation}
c_2^2(r) = \frac{1+\sqrt{1-4 \lambda}}{2 r} f(r) \frac{1- \lambda f''(r)}{r - \lambda f'(r)}
\end{equation}

The 2 as a lower index indicates that this calculation was performed for the helicity 2 mode of the graviton.

\begin{figure}[htb]
\begin{center}
\epsfxsize=2.6in\leavevmode\epsfbox{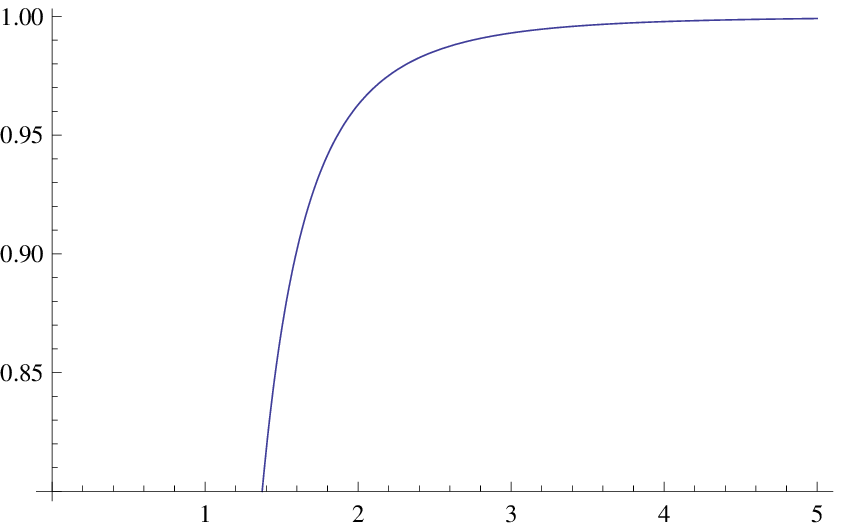}
\epsfxsize=2.6in\leavevmode\epsfbox{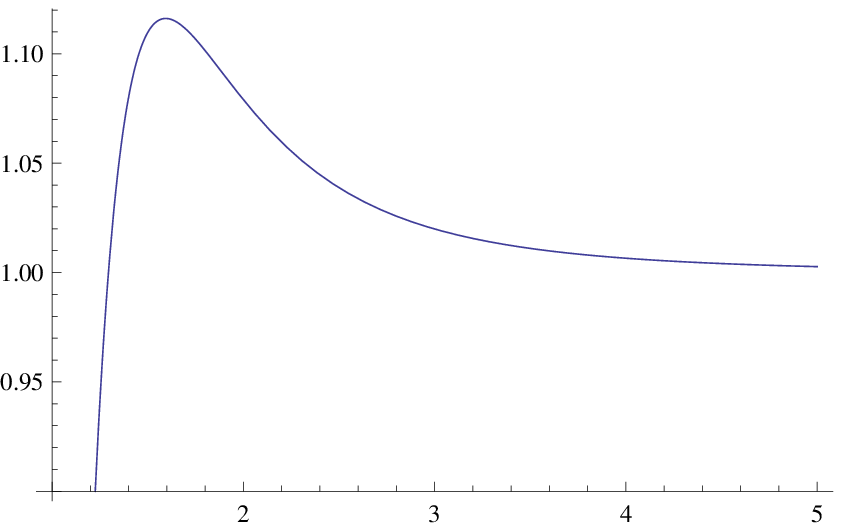}
\end{center}
\caption{\footnotesize{ Plots of $c^2_2(r)$.  (a) Plot for $\lambda=0.05<\frac{9}{100}$. (b) Plot for $\lambda=0.15>\frac{9}{100}$. Notice that $c^2_2>1$ for $r$ close enough to the boundary $r \rightarrow \infty$ when $\lambda>\frac{9}{100}$.}} \label{c2plot}
\end{figure}

By analyzing $c_2^2(r)$ we can see that it is such that $c_2^2<1$ for all $r$ for $\lambda<\frac{9}{100}$ and it is monotonous. There is no
turning point. For $\lambda>\frac{9}{100}$,  $c_2^2$ becomes greater than 1 in part of the range and develops a unique maximum where $c_2^2>1$. Therefore, causality would be broken at the boundary if
$\lambda>\frac{9}{100}$.

We can make an interesting comment at this point. If we look in detail at figure \ref{c2plot}(b) we see that $c^2>1$ for all values of r close enough to the boundary. This fact suggests that any problems encountered should be seen by looking at the UV behavior of our field theory.

We would now like to repeat this calculation for a different polarization. The results in \cite{ccp} indicate that a lower bound on $\lambda$
might be obtained by studying  a state with helicity 0 in the CFT. The symmetries are not enough in this case to select only one component of the
metric as our degree of freedom. If we pick a gauge where $h_{\mu 0}=0$, we still have to consider the dynamics of four components of the metric
perturbation. Namely, $h_{11}=h_{22}$, $h_{33}$, $h_{rr}$ and $h_{3r}$. Solving the equations of motion for these components is complicated in general, but luckily we are only  interested in the value of the speed of light in the
3-direction. Then, we can keep only terms of leading order in $\partial_0$ and $\partial_3$. For a plane wave solution $e^{i \omega x^0 - i k x^3}$,
the equations of motion become algebraic equations and can be solved. This yields:

\begin{eqnarray}
h_{3r} &=& 0\nonumber\\
h_{rr} &=& \frac{4 r(r- \lambda f'(r))}{-2 r^2 + 4 \lambda f(r)} h_{11}\\
h_{33} &=& \left(-2 +  \frac{6 (r -\lambda f'(r))^2}{3 r^2 + 2 \lambda f(r) - 8 r \lambda f'(r) + 4 \lambda^2 f'^2(r) + \lambda (r^2-2 \lambda f(r))f''(r)}   \right) h_{11}\nonumber
\end{eqnarray}

The remaining equation of motion imposes

\begin{equation}
\frac{\omega^2}{k^2} = c_0^2 = \frac{(1+\sqrt{1-4\lambda})f(r)(3 r^2 + 2 \lambda f(r) - 8 r \lambda f'(r) + 4 \lambda^2 f'^2(r) + \lambda (r^2-2 \lambda f(r))f''(r))}{6r(r^2-2\lambda f(r))(r -\lambda f'(r))}
\end{equation}

This is the expression for the speed of light in the 3-direction for the helicity 0 modes. If we again repeat the analysis of the curve given by
$c_0^2(r)$ as done for the helicity 2 mode we find a similar behavior\footnote{The corresponding plots look qualitatively the same as the ones presented in figure \ref{c2plot} and we omit them for brevity.}. The main difference is that in this case we observe acausal behavior (a maximum in $c^2_0$ with $c^2_0>1$) for
$\lambda<-\frac{7}{36}$. This corresponds to $\frac{a}{c}>\frac{3}{2}$, as expected. We therefore see that if we extend the analysis of
\cite{liu} to this case we match the bound obtained in \cite{ccp}.

Interestingly enough, we can also study the helicity 1 mode of the graviton. Although, this is not an optimal bound, we also obtain the expected results predicted by \nref{boundsac}. Namely,
$\frac{a}{c} < 2 \rightarrow \lambda > -\frac{3}{4}$. We
quote the speed of light in this case.

\begin{equation}
c_1^2 = \frac{(1 + \sqrt{1- 4 \lambda})(r- \lambda f'(r)) f(r)}{2 r (r^2 - 2 \lambda f(r))}
\end{equation}

There are two independent perturbations of the metric in this case after gauge fixing: $h_{31}$ and $h_{r1}$. The latter is set to zero by the
equations of motion, while the former is the physical degree of freedom.

We have seen that this analysis allows one to calculate that a consistent conformal field theory must have $\frac{1}{2} < \frac{a}{c} <\frac{3}{2}$ if bulk dynamics are to induce causal behavior at the boundary. This is the expected window for a CFT with $\mathcal{N}=1$. How do we get the stronger bound found for $\mathcal{N}=2$ theories in \cite{ccp,Shapere:2008zf}? At the level of the low energy four derivative action \nref{gaction2} $\mathcal{N}=1$ and $\mathcal{N}=2$ theories are equal except for the fact that $\mathcal{N}=2$ theories possess an additional $SU(2)$ gauge field in their dual gravitational description. If we are to find the narrower window $\frac{1}{2} < \frac{a}{c} < \frac{5}{4}$ we should look at the propagation of the helicity 0 mode of the $SU(2)$ gauge field, as suggested by \nref{boundsac}. We do this in the next subsection.

\subsection{Gauge bosons in the black hole background}
\label{gaugeconf}

Now we would like to obtain bounds \nref{a21} and \nref{a22} from considering the propagation of gauge bosons in a black hole background. The simplest way to do this calculation is to consider Einstein gravity with no higher derivative corrections and add Maxwell terms (including the correction proportional to $W$) as in \nref{gaction2}. We will comment on the full action \nref{gaction2} when we discuss shock waves further below. We will obtain the same bounds by using action \nref{gaction2} in this context.

 In our coordinates, the (Einstein) metric of the black hole takes the form

\begin{equation}
ds^2 = - \left(\frac{r^4-1}{r^2}\right) dt^2 + \left(\frac{r^2}{r^4-1}\right) dr^2 + r^2 dx^i dx^i
\end{equation}

We can now use the electromagnetic part of the action \nref{gaction2} in this background to obtain effective actions for the 0 and 1 helicity modes of the gauge field. Up to quadratic order in the electromagnetic perturbation it does not make a difference whether we have an abelian or non abelian gauge group.  For the helicity 1 mode we can set $A = A_1(r) e^{i k x^3 - i \omega x^0} dx^1$. The effective equation of motion for this mode at $|k|, \omega >> |\partial_r|$ is:

\begin{equation}
\left((-1 + r^4) (6 r^4 + a_2) k^2 +
  r^4 (-6 r^4 + a_2) \omega^2\right) A_1 =0
 \end{equation}

 Therefore,

 \begin{equation}
 c_{1A}^2 = \frac{(r^4 - 1) (6 r^4 + a_2)}{r^4 (6 r^4 - a_2)}
 \end{equation}

 The behavior of $c_{1a}^2$ is analogous to that of $c^2_i$ discussed before. We find acausal behavior for\footnote{This behavior has also been discussed in \cite{Ritz:2008kh}.} $a_2>3$. This matches exactly the expectations from \nref{a21}. We can now use the relations \nref{u1} and \nref{su2} to express this bound as a function of $\frac{a}{c}$. We find that if we consider a $U(1)$ gauge field this yields $\frac{a}{c}>0$ while and $SU(2)$ field yields $\frac{a}{c}>\frac{1}{2}$. These are the predicted results form the energy one point function analysis listed in \nref{boundsac}. This result does not improve the window coming from the gravitational excitations. In order to do that we have to consider the 0 helicity mode. In that case we can gauge fix $A_0=0$. The equations of motion impose $A_3=0$ and give us the following value for the speed of light squared:

 \begin{equation}
 c_{0A}^2 = \frac{( r^4-1) (6 r^4 - a_2)}{3 r^4 (2 r^4 + a_2)}
 \end{equation}

Once again, we observe a similar behavior, indicating problems with causality for $a_2<-\frac{3}{2}$. This time we see that while for the case of a $U(1)$ field this implies $\frac{a}{c}<\frac{3}{2}$, and no new bound is found, the presence of an $SU(2)$ gauge field imposes $\frac{a}{c}<\frac{5}{4}$, as predicted in \cite{ccp,Shapere:2008zf}. Therefore gravitational theories which are dual to $\mathcal{N}=2$ supersymmetric conformal field theories have a narrower causal window.

\subsection{Are all higher derivative gravities equal?}

Up to this point all seems well and in agreement between the field theory results and the higher derivative gravity calculation.  One small exercise one could attempt to understand whether all parameterizations of the 4 derivative gravity actions are equivalent is to reproduce the results in section \ref{gravbh} using action \nref{gaction2}. Here, the scene is complicated by the fact that the equations of motion for the gravitational perturbations include terms with up to four derivatives. We remind the reader that the absence of these terms is what makes Gauss-Bonnet gravity special. Although, this can be done, it involves dealing with the black hole solutions of other gravity theories. Instead of pursuing this road, we will study a different background in the next section, namely, the shock wave. We will explain in the next section several reasons why this is a good idea. In this setup, it will be easier to compare different actions.

At this point we should spoil the punch line and say that we won't see the same results when looking at classical solutions using actions \nref{gbaction} and \nref{gaction2}.

 One might be tempted to argue that there are other metric perturbations in the theory that might be responsible for the problem. Although this modes are problematic on their own, as being responsible for non unitarity in the quantum theory (unless we embed the model in a UV complete theory), they are dual to operators with different weight in the dual CFT. In other words, they are massive fields in the gravitational theory. The only point when they do become massless gravitons corresponds to $\frac{a}{c}=\infty$, when they \textit{cancel} the usual gravitons and the theory becomes topological Chern-Simons.  Then, except for $\frac{a}{c}=\infty$ we can always imagine imposing boundary conditions such that the only mode we excite is the one that couples to the energy-momentum tensor. See more on this in section \ref{sectw}.

We will see in the next section that one explanation for the different behavior in this case is that graviton scattering of a classical background involves higher energy point functions for a general 4 derivative gravity theory. It is only in Gauss-Bonnet gravity where the scattering of gravitons of a shock wave involves only the energy one point functions of the dual CFT. Remember that the gravitational actions of the type discussed in \cite{ccp} are engineered to reproduce the energy one point function. When higher point functions become relevant we need to consider higher derivative terms and we can't truncate the form of the action to \nref{gaction2}.

\section{Shock wave backgrounds}
\label{shock}

It seems like studying a black hole background is overkill for the type of problem we have been discussing. This amounts to putting the theory in a thermal bath. It is natural to
think that the simplest thing to consider is the interaction of gravitons with shock waves that propagate on top of AdS space-time. This matches
naturally with the observables considered in \cite{ccp}, where it was explained that this is the gravitational configuration dual to the energy one point function in the CFT. Let us first check that shock waves are indeed a solution to the equations of motion,
modified by higher derivative terms.

\subsection{Solutions to Gauss Bonnet gravity and linearity}

Although this is a trivial fact, we recall  that $AdS$ is a solution to Gauss Bonnet gravity. This can be checked by taking the zero mass limit
of the black hole solution discussed in \cite{liu}. We can also check it explicitly from the equations of motion. It is important to remember (and this is
why we are discussing this trivial fact) that the higher derivative correction changes the curvature of the AdS space. For a given value of the
cosmological constant, normalized such that the unperturbed AdS space has radius 1, the radius of the AdS space in Gauss-Bonnet gravity is given by:

\begin{equation}
R^2 = \frac{1}{2} \left(1+\sqrt{1- 4 \lambda}\right)
\end{equation}

This reduces to 1 as $\lambda \rightarrow 0$ and shows explicitly that there is no AdS solution for $\lambda>\frac{1}{4}$.  We choose to
parameterize the AdS metric with Poincare coordinates as\footnote{The z coordinate corresponds to $\frac{1}{r}$.}:

\begin{equation}
ds^2_{AdS} =\frac{1}{2} \left(1+\sqrt{1- 4 \lambda}\right) \frac{dz^2 + d\vec x^2}{z^2}
\end{equation}

We can now add a shock wave propagating on top of this metric. We define coordinates $x^\pm = x^0 \pm x^3, x=x^1, y=x^2$ and take the following
form for a general shock wave metric

\begin{equation}
ds^2_{Shock} = ds^2_{AdS} + \delta(x^+) w(x,y,z) dx^{+2}\label{metricsh}
\end{equation}

While $\delta(x^-)$ is in principle an arbitrary function we will consider it to be a localized delta function.
By plugging \nref{metricsh} into the equations of motion for Gauss-Bonnet gravity we find that this is always a solution provided

\begin{equation}
4 w - z \partial_z w - z^2 \left(\partial_z^2 w + \partial_x^2 w + \partial_y^2 w \right)=0
\end{equation}

This is the same condition we find in usual Einstein gravity for propagation of a shock wave on $AdS$. We can see that this equation is linear, so we can superimpose as many solutions as we want. This result means that, classically, shock waves do not
interact with each other.  This result was already briefly discussed in \cite{ccp}, in the context of stringy corrections to the energy correlation functions. The fact that the shock wave solution receives no corrections from higher derivative terms was discussed at length in \cite{Horowitz:1999gf}.

Some interesting types of solutions to this equation are:

\begin{eqnarray}
w^A &=& \alpha\frac{1}{2} \left(1+\sqrt{1- 4 \lambda}\right) z^2\label{wa}\\
w^B &=& \alpha \frac{1}{2} \left(1+\sqrt{1- 4 \lambda}\right)\frac{1}{z^2}\label{wb}\\
w^C &=& \alpha  \frac{1}{2} \left(1+\sqrt{1- 4 \lambda}\right)\frac{z^2}{(z^2+x^2+y^2)^3}\label{wc}
\end{eqnarray}

\noindent where $\alpha$ is just an arbitrary normalization. Notice that these are exact solutions of Gauss Bonnet gravity, therefore this is
not an expansion in $\alpha$.

While solutions $A$ \nref{wa} and $C$ \nref{wc} represent interesting physical situations, $B$ \nref{wb} is just a coordinate redefinition. We will see this explicitly when
we study the propagation of perturbations on top of $AdS$ space with shock waves of these types.

\subsection{From the Black Hole to the shock wave}

One way to make contact between the black hole solution and the shock wave is to realize that the shock wave solution can be obtained by
boosting a black hole solution while keeping its energy $E=\frac{m}{\sqrt{1-v^2}}$ constant \cite{sexl,thooft,Horowitz:1999gf}. In our particular case, we are studying a black brane, which is invariant under translations in the boundary coordinates. It is possible to check that by boosting this solution we obtain shock wave $w^A$ \nref{wa}, which retains the symmetries preserved by the boost. This solution corresponds to the deformation of the background caused by some source deep inside $AdS$. In this case the normalization constant $\alpha$ is proportional to the energy density (which we keep constant) of the black brane and it is positive if the solution has a positive mass. 

We can now calculate the effective linearized equation of motion for gravitons with helicity 2 (as in \cite{liu}) in this background. The
resulting equation of motion for the perturbation $h(x^+,x^-,z) dx\, dy$ is

\begin{equation}
\frac{3}{z} \partial_z h  - \partial^2_z h + 4 \partial_+ \partial_- h + 4 \alpha \delta(x^+) \left[1-16 g(\lambda)\right] z^4 \partial^2_- h =0\label{shockeom1}
\end{equation}

\noindent with $g(\lambda)=\frac{\lambda}{1-4\lambda + \sqrt{1-4\lambda}}$.

In the limit of very high energies we can consider a wave packet that moves with definite momentum. By integrating over the discontinuity at
$x^+=0$ we can extend a solution $h_<$  at $x^+<0$ to the other side of the shock wave ($x^+>0$) as

\begin{equation}
h_> = e^{-i P_- \alpha \left[1-16 g(\lambda)\right] z^4}h_<\label{disc1}
\end{equation}

\noindent where we used $P_- = - i \partial_-$. From this we can read there is a shift $\Delta x^- = \alpha\left[1-16 g(\lambda)\right] z^4$
after the collision. Also, we can act on \nref{disc1} with $P_z = -i \partial_z$. We obtain the shift in the momentum in the z direction as:

\begin{equation}
P_z^> = P_z^< - 4 P_- \alpha  \left[1-16 g(\lambda)\right] z^3
\end{equation}

Notice that for a particle going inside $AdS$, $P_z>0$. Therefore, if we want our particle to come back to the boundary we need $\left[1-16
g(\lambda)\right]<0$. This comes about from the fact that $P_-<0$ and $\alpha>0$ if our original black hole had a positive mass. But if this is
the case, then $\Delta x^- <0$. Notice that the shock wave vanishes on the boundary  \nref{wa}. Therefore, the shock wave does not affect the light cone of particles that move on the boundary of $AdS$ ($z=0$). The end result is that particles that come back to the boundary end up landing outside the light cone.

If we require that $1-16 g(\lambda)>0$, we recover the result from \cite{liu} $\lambda<\frac{9}{100}$. If the bound is satisfied, particles that go into the
bulk of $AdS$ do not return to the boundary as found in \cite{liu} (also, look at the form of $c^2$ in figure \ref{c2plot}(a)), and no problem is encountered.

\subsection{Shock waves that change the boundary metric}

Let us now look at cases where we actually insert a perturbation at the boundary. This corresponds to non-normalizable shock waves of the type
$B$ \nref{wb} and $C$ \nref{wc}. We are interested in this possibility as we want to make a connection to the calculation of correlation functions in the boundary
theory.

The first thing we can check is that, because $B$ corresponds to a change of coordinates, the equation of motion of a gravitational perturbation
will not depend on $\lambda$ or the point $z$ inside $AdS$ where the collision occurs. In accordance with our expectations, the equation of
motion for the perturbation is, in this case:

\begin{equation}
\frac{3}{z} \partial_z h  - \partial^2_z h + 4 \partial_+ \partial_- h + 4 \alpha \delta(x^+)\partial^2_- h =0
\end{equation}

Therefore, the discontinuity of the light cone is the same for all propagating modes and there is no conflict with causality. Also, particles
can't bounce back to the boundary.

The shock wave given by $w^C$ is more interesting. It corresponds to the insertion of a stress energy tensor over a light like line at the
boundary. In this case

\begin{equation}
\frac{3}{z} \partial_z h  - \partial^2_z h + 4 \partial_+ \partial_- h +\frac{ 4 \alpha \delta(x^+) z^4}{(z^2+x^2+y^2)^3}  \left(\left[1-16 g(\lambda)\right] + 96 g(\lambda) \frac{z^2(x^2+y^2)}{(z^2+x^2+y^2)^2}\right) \partial^2_- h =0 \label{shockeom2}
\end{equation}

The situation is a little bit more complicated in this case as there is a dependence on where we put our particle in the transverse space
$(x,y)$. The particle trajectory is shifted by

\begin{equation}\label{timeshift}
\Delta x^- =\frac{ \alpha z^4}{(z^2+x^2+y^2)^3}  \left(\left[1-16 g(\lambda)\right] + 96 g(\lambda) \frac{z^2(x^2+y^2)}{(z^2+x^2+y^2)^2}\right)
\end{equation}

If we study this problem for a particle at $x=y=0$ we find

\begin{equation}\label{ts0}
\Delta x^- =\frac{ \alpha}{z^2} \left[1-16 g(\lambda)\right]
\end{equation}

and

\begin{equation}
P_z^> = P_z^< + 2 P_- \frac{\alpha}{z^3}  \left[1-16 g(\lambda)\right]
\end{equation}

This implies that we need $\alpha \left[1-16 g(\lambda)\right]>0$ for our particle to come back to the boundary. In order to see whether we have
problems with causality we need to study the light cone of the boundary theory. In this particular case the metric of the boundary theory is
affected, so we need to be careful.

Because the shock wave at the boundary does not depend on $\lambda$ we expect the light cone to be changed in a $\lambda$
independent manner. It is clear from the equation of motion \nref{shockeom2} that the shock wave behaves as $ \sim \delta(x^+) \delta(x,y)$ at
the boundary. We need to check what the overall multiplicative factor is. In order to check this we will integrate the shock wave over $x$ and
$y$ at a fixed $z$ to obtain $w_{eff}(z)$ and then take the $z \rightarrow 0$ limit. Therefore,

\begin{eqnarray}
w_{eff}(z) &=& \int dxdy \frac{ \alpha z^4}{(z^2+x^2+y^2)^3}  \left(\left[1-16 g(\lambda)\right] + 96 g(\lambda)
\frac{z^2(x^2+y^2)}{(z^2+x^2+y^2)^2}\right)\\
&=& \frac{\alpha}{4}\left[1-16 g(\lambda)\right] +\frac{\alpha}{24} 96 g(\lambda) = \frac{\alpha}{4}
\end{eqnarray}

We can see that this result is independent of $z$ and $\lambda$. Therefore the light cone shift at the boundary is:

\begin{equation}
\Delta x^-\big|_{z=0} \sim \alpha \delta(x,y)
\end{equation}

This infinite shift at $x=y=0$ is a consequence of considering a localized discontinuous perturbation of the boundary and could be regularized
by a smooth deformation of the metric. If the light cone shifted towards $x^-<0$ by an infinite amount, then every event with $x^+>0$ would be inside the future cone of our null geodesic. Therefore, we would like to consider the case in which $\Delta x^-\big|_{z=0} >0$ so we can have a possible
violation of causality. The schematic picture of the situation is the following

\begin{figure}[htb]
\begin{center}
\epsfxsize=2.6in\leavevmode\epsfbox{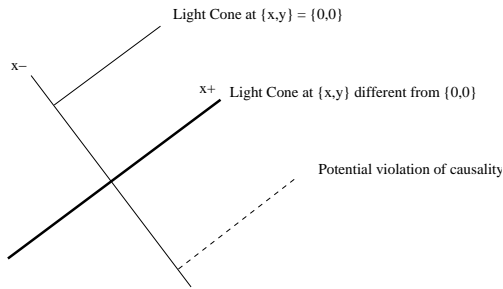}
\end{center}
\caption{ \footnotesize{Light cone structure of our boundary space time. The thick line represents the unchanged null geodesic away from the shock wave in transverse space $\{x,y\} \neq \{0,0\}$. The thin line is the displaced geodesic at the position of the shock wave $\{x,y\}=\{0,0\}$. Notice that if we have a $\delta$ function type shock wave this geodesic sits at $x^-=\infty$. If we find that $\Delta x^- < 0$ for the metric perturbation (dashed line), we will encounter a causality problem.}} \label{lcone}
\end{figure}

Having the boundary light cone shift in the positive $x^-$ direction  implies $\alpha>0$. Therefore we need $1-16 g(\lambda)>0$ for the particle to come back to the boundary and,
according to (\ref{ts0}) we don't have a violation of causality.

We can now consider a particle at $x, y \neq 0$. In that case and for $z \ll x,y$  we can write the shift in $x^-$ as

\begin{equation}\label{timeshift2}
\Delta x^- =\frac{ \alpha z^4}{(x^2+y^2)^3}  \left[1-16 g(\lambda)\right]
\end{equation}

This time

\begin{equation}
P_z^> = P_z^< - 4 P_- \alpha  \left[1-16 g(\lambda)\right] \frac{z^3}{(x^2+y^2)^3}
\end{equation}

Now this case is almost identical to the one discussed for the shock wave coming from the black hole solution. We will observe a violation in causality
for $1-16 g(\lambda)<0$. Notice that in this case there is also a small deflection in the $x,y$ directions. This effect is however
suppressed as $\frac{z}{\sqrt{x^2+y^2}}$ with respect to the deflection in $z$.

\subsection{General remarks about results for the shock wave background}

The results of the last two subsections make it clear that the causality problems found from the gravity theory are directly connected with the value of the energy one point function. The calculation of the energy one point function is given by the scattering of a shock wave and we have shown that here is where the problem lies. Notice also that the equations of motion for the gravitational perturbation \nref{shockeom1} and \nref{shockeom2} are exactly linear in $\alpha$. This indicates that only the 3 point function involving two gravitons and a shock wave play a part in this discussion. As explained in \cite{ccp}, this means that only the energy one point function is of relevance here.

Let us explain in more detail why it is only the three point function that it is involved. The equations of motion studied, \nref{shockeom1} and \nref{shockeom2},  are of the form generic form:

\begin{equation}
(\Delta - \delta_{sw} \partial_-^2) h =0
\end{equation}

\noindent where $\Delta$ is the usual differential operator in $AdS$ for the propagation of a perturbation and $\delta_{sw}$ represents the shock wave. Therefore, the classical (tree level) propagator is:

\begin{equation}
P \sim \frac{1}{\Delta - \delta_{sw} \partial_-^2} = \frac{1}{\Delta} \sum_n \left(\frac{\delta_{sw} \partial_-^2}{\Delta}\right)^n
\end{equation}

This exponentiation corresponds to the following scattering picture, where, as we can see, only the three point function is involved.

\begin{figure}[htb]
\begin{center}
\epsfxsize=5in\leavevmode\epsfbox{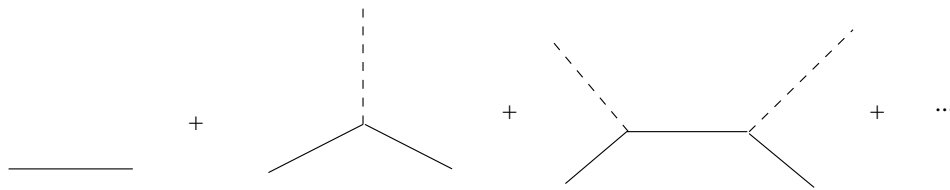}
\end{center}
\caption{\footnotesize{Only the three point function contributes to the exponentiation given by the classical solution. Solid lines represent the perturbation while the dashed lines represents the shock wave.}} \label{scat}
\end{figure}

Finally let us add that, because we have found a problem for scattering that occurs arbitrarily near the boundary  ($z \sim 0$) in \nref{timeshift2} , the calculation makes manifest that any potential problem reflects the UV properties of the field theory. We could have reached a similar conclusion by following the analysis of \cite{liu} and avoiding the argument using the maximum in the effective potential. In this case, calculating the exact classical trajectories would have shown the same problem. All this is clear from the fact, already noted in section \ref{gravbh}, that figure \ref{c2plot} shows that $c^2>1$ for small values of $z$. The upshot of this discussion is that the positivity of energy condition explained in \cite{ccp} must apply, not only to CFTs, but to any UV complete QFT (i.e. field theories that are asymptotically free or have a UV fixed point). 
Needless to say, we could repeat the same analysis for the other graviton polarizations and obtain similar results.

\subsection{The shock wave background for $W^2$ higher derivative gravity}
\label{sectw}

In this part we would like to study the problem of scattering perturbations of a shock wave by using a gravity action of the form \nref{gaction2}. We will only consider a shock wave of the form $w^A$ in \nref{wa}. This solution is more symmetrical and further detail won't be necessary. 

First let us briefly study the problem for gauge boson perturbations so we can be sure that what was discussed in section \ref{gaugeconf} can be extended with no problems to the full action \nref{gaction2}. The argument is fast and simple. The shock wave remains a solution of the full equations of motion including the higher derivative terms. Because, once we have a background, all we need is the Maxwell part of the action, the calculation is the same we would have in Einstein gravity or GB gravity. We spare the reader the details in this case and just note that studying helicity 1 and 0 modes we recover the results \nref{a21} and \nref{a22}.

In the case of gravitational perturbations the story is more complicated. The action \nref{gaction2} yields fourth order equations and, as we shall see, also involves a vertex involving two shock waves. If we are only interested in the high momentum limit we can neglect the Einstein term and just study the $W^2$ contribution. We see right away that, because the shock wave solution is independent of $t_2$, the equations of motion will not depend on $t_2$ at all. This already shows that it will not be possible to match the bounds on $t_2$ listed in \nref{t21},\nref{t22} and \nref{t23}. Let us understand where the problem comes from. The equations of motion in the $\partial_-,\partial_+ \rightarrow \infty$ limit for a helicity 2 component of the metric $h=h_{12}$ are:

\begin{equation}\label{eom4}
\left(\partial_+^2\partial_-^2 + z^4 \delta'(x^+) \partial_-^3 + 2 z^4 \delta(x^+) \partial_+ \partial_-^4 + z^8 \delta(x^+)^2 \partial_-^2\right) h = 0
\end{equation}

This can be rewritten as:

\begin{equation}
\left(\partial_+\partial_- + z^4 \delta(x^+) \partial_-^2\right)^2  h = 0
\end{equation}

This expression makes manifest that there are no problems of causality of the type discussed before. The shift in $x^-$ is always positive. Notice that in this limit the metric has two independent spin 2 degrees of freedom. If we had kept the contributions from subleading terms in $\partial_-,\partial_+$, we would see that one of these modes becomes massive while the other is the usual graviton. Any problems induced by a possible tachyonic mass are visible at low momentum and, in any case, represent a different phenomenon. As we explained, the only point where both modes are massless is the Chern-Simons theory ($\frac{a}{c} \rightarrow \infty$). The limit of infinite mass, $\frac{a}{c} \rightarrow 1$, is such that the theory is tachyonic for $\frac{a}{c}$ slightly below 1 and has $m^2>0$ for $\frac{a}{c}$ slightly above 1. This makes manifest that these problems are unrelated with the positivity of energy in the dual CFT.

Let us look at these statements in more detail. The equation of motion for a perturbation of the metric of helicity 2, $h(z) e^{i k x^3 - i \omega x^0} dx^1 dx^2$, over $AdS$ given by the action \nref{gaction2} is

\begin{eqnarray}
0 & = & \left[21 z  (\omega^2-k^2)   + 8\beta z^3 (\omega^2-k^2) (\omega^2-k^2-\frac{1}{z^2}) \right] h(z) +\nonumber\\
& & \left[-9 + 32 \beta z^2 (\omega^2-k^2)\right] h'(z) +  \left[3 z + 16 \beta z (1- 2 z^2(\omega^2-k^2))\right] h''(z) - \nonumber\\
& &   16 \beta z^2 h'''(z)+ 8 \beta z^3 h''''(z)
\end{eqnarray}

\noindent where $\beta = \frac{t_2}{48-8 t_2} = \frac{1}{8}\left(\frac{c}{a} -1\right)$.

If we solve this equation asymptotically near the boundary, $z\rightarrow 0$ with an ansatz of the form $z^\Delta$ we find the following solutions:

\begin{equation}
\Delta_-^1 = 0 \quad\quad \Delta_+^1 = 4 \quad\quad \Delta_-^2 = 2 - \sqrt{1-\frac{3}{8\beta}} \quad\quad \Delta_+^2 = 2 + \sqrt{1-\frac{3}{8\beta}}
\end{equation}

Therefore, $\Delta^{1,2}_+$ are the conformal dimensions of the spin 2 operators dual to the degrees of freedom contained in the metric. While $\Delta_+^1$ corresponds to the usual stress-energy tensor the other gravitational mode is dual to a different spin 2 operator of weight $\Delta_+^2$. As promised, one can see that the second degree of freedom becomes \textit{another} graviton for $\beta=-\frac{1}{8}$ ($\frac{a}{c} = \infty$) where they cancel and no degree of freedom remains. Also, the massive mode is such that $m^2 \sim \left(\Delta^2_+\right)^2$ for $\Delta^2_+ \rightarrow \infty$. A small value of $\beta$ yields $m^2 \rightarrow \infty$ for $\beta<0$ and $m^2 \rightarrow -\infty$ for $\beta>0$. We therefore confirm that the peculiarities of the second spin 2 mode are not directly related to the causality problems discussed above. 

Going back to our discussion of the large momentum limit, the main reason why we expect things to be different from GB gravity is that the equations of motion \nref{eom4} include vertices that involve two shock waves ($\delta^2$). This goes beyond the graviton 3 point function where we can trust our general description by the action \nref{gaction2}. The main point is that GB gravity is special in the sense that it only involves the 3 point function when we exponentiate tree level diagrams, while a generic $2n$ derivative theory involves up to $2+n$ graviton vertices.

One might add that the reason why no $\delta^2$ contributions appear in \nref{shockeom1} is a combination of 2 derivative equations of motion with Lorentz symmetry. As we can see from \nref{metricsh}, the shock wave $w$ has two $+$ lower indices. The only way to obtain a Lorentz invariant expression is to be able to construct a term with as many lower $-$ indices. If we want a $\delta^2$ term we need to have $\partial^4_-$ terms available. As these can't appear in Gauss-Bonnet gravity, only the three particle vertex can contribute. The same reasoning proves that any $2n$ derivative theory of gravity has at most $\delta^{n}$ terms.

The end result of this discussion is that, while actions \nref{gaction2} and \nref{gbaction} are engineered to represent the same energy one point function in their dual theories (they yield the same value for $\frac{a}{c}$, that is), when we consider classical solutions the exponentiation procedure is different and one theory involves higher point vertices than the other. Therefore, one must be careful about what consistency results are expected from arbitrary higher derivative theories of gravity.

\section{Positivity of energy in any CFT from field theory arguments}
\label{positivity}

In this section we would like to add a short argument explaining why, besides expectations coming from the experimental collider setup, the energy one point function should be positive in any CFT. The reader should take the following arguments as a \textit{physicist's proof} as opposed to a formal proof (as in Axiomatic Quantum Field Theory, say).

The first thing that should be said is that results in \cite{ccp} show that, using conformal transformations, the positivity of the energy operator  \nref{eop} is equivalent to the positivity of the following energy operator written in new coordinates:

\begin{equation}\label{eop2}
{\cal E}(\vec y) = \int dy^- T_{--}\left(y^-, y^+, \vec y\right)
\end{equation}

\noindent where the flat space metric is $ds^2 = - dy^+dy^- + d\vec y^2$. Therefore, what we are trying to prove is a version of the averaged null energy condition for CFTs in flat space.

It was pointed out in \cite{ccp} that proving the positivity of \nref{eop2} is an easy task for free  theories. Using the creation and annihilation operators one can show that the integral in $y^-$ precisely takes care of the terms in $T_{--}$ responsible for negative energy densities \cite{Klinkhammer:1991ki}.  We are, therefore, interested in writing the argument in a general language common to all CFTs that we can apply to strongly coupled theories. Furthermore, we want to use minimum structure, as the argument should apply to all CFTs. We are led to study the operator product expansions (OPEs) of energy momentum tensors.  Although this structure is present in every theory, the OPEs in $d>2$ are not as thoroughly determined as their $d=2$ counterparts. An example of this indeterminacy are Schwinger terms. We will, however, make use of all information available to constraint our results.

The general strategy will be the following. We will consider non local operators similar to \nref{eop2}. Then we consider the theory with euclidean time $y^+$ and we regard $y^-$ as a space variable. If we can find one operator such that its OPE with itself yields \nref{eop2} at leading order in $y^+$, then the standard argument of positivity of norms\footnote{See \cite{Grinstein:2008qk} for an interesting recent discussion of this old argument.} proves the positivity of our energy operator \nref{eop2}.

Schematically, if
\begin{equation}\label{ope1}
\mathcal{O}(0) \mathcal{O}(y^+) \sim \frac{\gamma}{y^{+n}} \int dy^- T_{--}(0) + \ldots
\end{equation}

\noindent then $\int dy^- T_{--}(0)$ is a positive operator, provided $\gamma>0$. This means that any $N$ point function of the form

\begin{equation}
\langle \alpha| {\cal E}(\vec y_1) \ldots {\cal E}(\vec y_N)|\alpha\rangle >0
\end{equation}

\noindent for any state $|\alpha\rangle$. 

Let us find candidate operators $\mathcal{O}$.  Equation \nref{ope1} plus invariance under the full conformal group imply

\begin{eqnarray}
2 \Delta\left(\mathcal{O}\right) &=& n + 3\\
2 S\left(\mathcal{O}\right) &=& -n + 1
\end{eqnarray}

\noindent where  $\Delta\left(\mathcal{O}\right)$ is the conformal weight  and $S\left(\mathcal{O}\right)$ is the Lorentz spin of $\mathcal{O}$ in the $+-$ plane. This implies

\begin{equation}
\Delta\left(\mathcal{O}\right) + S\left(\mathcal{O}\right) = 2
\end{equation}

We want to build an OPE from an operator $\mathcal{O}$ available in any theory. Natural candidates are $\int dy^- T_{\mu\nu}$, which have $\Delta=3$. Then, we need $S=-1$, which fixes $\mu\nu=+-$.  Our task is now to compute

\begin{equation}\label{ope2}
\int dy^- T_{-+}(0)\int dy^- T_{-+}(y^+) \sim \ldots + \frac{\gamma}{y^{+3}} \int dy^- T_{--}(0) + \ldots
\end{equation}

Our argument will be complete if we show that: a) the first group of $\ldots$ representing terms more divergent than $\frac{1}{y^{+3}}$ is  actually zero; b) $\gamma>0$.

More divergent contributions would be of the form $\frac{1}{y^{+k}} \mathcal{O}'_{\Delta',S'}$. Symmetry under the conformal group implies $\Delta' + k  = 6$ and $S'-k = -2$. Then,

\begin{equation}
k > 3 \quad\quad \longrightarrow \quad\quad \Delta' < 2 + S' \quad \textrm{and} \quad S'>1
\end{equation}

If we assume that only integrals of local operators can appear, then this contribution would violate the unitarity bound \cite{Mack:1975je}. It is interesting to notice that some local operators appear in the local OPE $T_{+-} T_{+-}$ but these contributions vanish when we integrate over $y^-$. This explains why local densities can be negative while \nref{eop2} is positive and is reminiscent of the situation in the free theory, discussed in \cite{Klinkhammer:1991ki}.

Of course, non local operators can appear in the OPE in principle (as they appear in a similar OPE in \cite{ccp}). We are not aware of a similar bound for this class of operators. If they do satisfy a similar bound, we know the leading term will be $\int dy^- T_{--}$. In that case, a) is true.

We can now show that b) is true by integrating \nref{ope2} over the transverse coordinates $\vec y$. Then, the left hand side of \nref{ope2} is still positive and in the right hand side we have $\sim\frac{\gamma}{y^{+3}} P^+$, where $P^+$ is the total momentum $P^+ \sim \int dy^- d\vec y^{\,2} T_{--} > 0$. Therefore $\gamma$ should be positive. We have checked this explicitly by using the 3 point functions in \cite{Osborn:1993cr} for the supersymmetric case and using the final result $\frac{3}{2} c > a > \frac{1}{2} c$ \cite{ccp}. In the explicit calculation, with  $\frac{3}{2} c > a > \frac{1}{2} c$, it is the case that $\gamma>0$, but the general argument does not rule out $\gamma=0$.

\section{Discussion}
\label{disc}

In this work we have discussed the connection between the constraints imposed by causality of the boundary theory on bulk dynamics and the positivity of the energy one point function in the field theory. Considering different propagating modes in the bulk leads to different causality constraints and there is a precise match between these bounds and the ones expected from the field theory analysis in \cite{ccp}. In particular we have reproduced the expected bounds for $\mathcal{N}=1$ and $\mathcal{N}=2$ supersymmetric CFTs by using gravitational and gauge boson perturbations in the bulk.

These calculations were performed for the black hole background and the shock wave background in higher derivative gravity. By performing the calculation in the shock wave background, several features of the problem become manifest. In the first place, this is the natural setup to perform the calculation. The AdS/CFT dictionary establishes that the shock wave background is the correct dual configuration to compute the energy one point function at zero temperature. Therefore, the connection between the energy one point function and the causality problem becomes evident. It is the graviton three point function (which upon integration yields the energy one point function) that is responsible for the scattering of the bulk modes outside the boundary theory light cone. Furthermore, it becomes evident that this is connected to properties of the theory at zero temperature.

This description of the problem also makes it clear that  causality violation reflects the UV properties of the dual CFT as the interaction can occur in the asymptotic region close to the boundary. This means that any UV complete QFT which flows to  fixed point at high energies will encounter the same problems, even when the black hole solutions discussed in \cite{liu} change dramatically. Therefore the positivity of energy must be a statement valid for all UV complete QFTs. While this statement is obvious for asymptotically free theories, as discussed in \cite{ccp,Klinkhammer:1991ki}, it is certainly non trivial for theories that flow to strongly coupled fixed points. One interesting point is that the argument can be reversed and used as a test of UV completeness for a candidate theory.  

It is also important to add that it is rather miraculous that the classical calculation in GB gravity yields the correct bounds. As was discussed in this work, different parameterizations of the 4 derivative gravity action that yield the same energy one point function do not show evident causality problems for propagation of perturbations on classical backgrounds. The case of a $W^2$ correction to the Einstein theory was studied in detail here. The small miracle of GB gravity comes about in the following way. When there is a problem with the three point function, one can observe scattering of perturbations only infinitesimally outside the light cone. This is hard to pick up. When one does a classical calculation the result from $n$ point functions is exponentiated. It is a special property of GB gravity that only the three point function contributes to this exponentiation. Therefore, the effect becomes amplified and can be observed macroscopically. A general gravity theory will have other contributions and the exponentiation can obscure the microscopic result. This is specially important to remember when using effective actions of the type \nref{gbaction} or \nref{gaction2} as they are built to reproduce the three point function exactly but can't be trusted if one is interested in higher point functions. The GB action is special in the sense that the classical calculation isolates the effect of the three point function.

Furthermore, it would be interesting to understand contributions of higher order terms in the gravity action. In particular, non supersymmetric theories require the presence of $R^3$ terms in the actions. It would be of interest to see if there is a classical gravity calculation that can reproduce the bounds proposed in \cite{ccp} for this case. 

Finally, we presented an argument in favor of the positivity of energy condition in a CFT. This result implies that the operator $\int dy^- T_{--}$ should be positive in any UV complete QFT (including strongly coupled and non supersymmetric theories). It would be interesting to have a more rigorous version of this proof. In particular, it would be nice to have a more detailed analysis of the contributions of non local operators.

\section*{Note}
While this work was being finalized we learned of results in \cite{Buchel:2009tt} which partially overlap with our results.

\section*{Acknowledgments}

I would like to thank Juan Maldacena especially for countless discussions on the subject and his constant encouragement. I would also like to thank Rob Myers for comments on an earlier version of the draft. It is my pleasure to also acknowledge interesting discussions with  Allan Adams, Nima Arkani-Hamed, Tolya Dymarsky, Jose Edelstein, Chris Herzog and especially John McGreevy. Finally I would like to thank the hospitality of KITP, where this work was finished. This  
research was  supported in part by the National Science Foundation  
under Grant No. PHY05-51164.

\end{document}